\documentclass[12pt]{article}
\begin{document}
\newcommand{\mpl}{M_{\mathrm{Pl}}}
\setlength{\baselineskip}{18pt}
\begin{titlepage}
%\begin{flushright}
%KOBE-TH-99-06 \\
%\end{flushright}

\vspace{2.5cm}

\centerline{{\LARGE\bf A Simple Proof of the 
Non-Renormalization }}
\vspace{5mm}
\centerline{{\LARGE\bf of the Chern-Simons Coupling}}
\vspace{1cm}

\centerline{Makoto Sakamoto$^{(a)}$
\footnote{e-mail : sakamoto@phys.sci.kobe-u.ac.jp}
and Hiroyuki Yamashita$^{(b)}$ \footnote{e-mail : 
hy257@phys.sci.kobe-ac.jp}
}
\vspace{1cm}
\centerline{$^{(a)}${\it Department of Physics, Kobe University,
Rokkodai, Nada, Kobe 657-8501, Japan}}
\centerline{$^{(b)}${\it Graduate School of Science and Technology, 
Kobe University,}}
\centerline{\it Rokkodai, Nada, Kobe 657-8501, Japan}
%
%
%   Abstract
%
\vspace{2cm}
\centerline{\large\bf Abstract}
\vspace{0.5cm}
We give a very simple proof that the renormalization of the
Chern-Simons coupling in the Wilsonian effective action
is exhausted at one-loop.
Our proof can apply to arbitrary
2+1-dimensional abelian as well as nonabelian gauge theories
without a bare Chern-Simons coupling,
including any non-renormalizable interactions and non-minimal couplings.
Our proof reveals that small (but not large)
gauge invariance is enough to ensure the absence of higher order
corrections.
\end{titlepage}

%
% Body
%

\section{Introduction}

Various powerful tools have been developed recently 
in understanding
the dynamics of supersymmetric gauge theories (for a review, 
see \cite{review-of-susy}).  Holomorphy is one of such tools, and 
Seiberg \cite{holomorphy} used it to greatly simplify the original proof
of the non-renormalization theorem \cite{gsr} and succeeded to find
the exact vacuum structures of various supersymmetric models.  
Although holomorphy is inherent in
supersymmetric theories, some of the techniques developed in
supersymmetric theories seem to be applied to 
non-supersymmetric ones.  
Nevertheless, very few are known about examples of such
applications.  The purpose of this letter is to present an example that
techniques developed in supersymmetric theories are applied to
non-supersymmetric ones to obtain (perturbative or non-perturbative)
exact results.

We consider a general (non-supersymmetric) gauge theory
in $2+1$-dimensions and give a new simple proof that
the renormalization of the 
Chern-Simons coupling is exhausted at one-loop 
by using techniques
developed in supersymmetric theories. 
Our proof makes it clear why the one-loop correction to
the Chern-Simons coupling is so special.
Most of our 
results have already been known but some of them are new.

Gauge theories in $2+1$-dimensions have a special feature to allow
a Chern-Simons term \cite{cs} (for a review, see \cite{review-of-cs})
\begin{equation}
  \int \!\! d^{3}x \ {\cal L}_{CS}
     = \int \!\! d^{3}x \ \kappa \:\epsilon^{\mu\nu\lambda}
        {\rm tr} \left( A_{\mu}\partial_{\nu}A_{\lambda}
        + \frac{2}{3} A_{\mu}A_{\nu}A_{\lambda} \right)\ .
  \label{cs-term}
\end{equation}
This term can be generated in one-loop order of perturbation theory
\cite{cs,one-loop1,one-loop2,one-loop3}
but the two-loop correction is shown 
to vanish \cite{two-loop1,two-loop2}.  
For abelian theories, 
there are no further corrections at
higher loops \cite{so,coleman-hill,matsuyama}.  
It is widely believed that
this is true even for nonabelian theories, although no rigorous proof
has been given.
This expectation is based on the topological nature of
the Chern-Simons term.
In nonabelian theories, this term is not
invariant under large gauge transformations, and consequently its
coefficient must be quantized to obtain consistent quantum theories.
If there were further corrections at higher loops, 
they would necessarily
spoil the quantized nature of the coupling.  
Another circumstantial evidence of the non-renormalization is 
that the $\beta$-function of the
Chern-Simons coupling vanishes in all orders of perturbation theory
\cite{b=0-1,b=0-2}, although this does not mean 
that there are no finite corrections to it.

The above topological reasoning to the non-renormalization of the
Chern-Simons coupling is, however, somewhat mysterious because
in perturbation theory we would not {\it a priori}
expect to \lq\lq know" anything about non-perturbative large
gauge transformations and
because the argument cannot apply to abelian theories.  
Our proof given in the next section can apply to both abelian
and nonabelian theories and relies only on gauge invariance under
{\it small} gauge transformations.  
Furthermore, our proof can apply to a
wider class of gauge theories than those discussed by Coleman and
Hill \cite{coleman-hill}.  
The authors proved the non-renormalization theorem
of the Chern-Simons coupling in a class of minimally coupled gauge
theories, in which gauge interactions are introduced by replacing
spacetime derivatives by covariant ones,
though their proof can allow non-renormalizable interactions.  
Gauge theories we consider
are more general and are not restricted to minimally
coupled ones.

The paper is organized as follows: In Section 2, some background of
recent developments in supersymmetric theories is explained and
then a very simple proof of the non-renormalization theorem is given.
In Section 3, several remarks are made.

\section{A Proof}

We consider a general $2+1$-dimensional gauge theory involving scalar
fields $\Phi_i$, spinor fields $\Psi_a$ and the gauge field $A_\mu$.  
The general gauge-invariant action we start with is of the form
%
%
%\begin{eqnarray}
%S 
%  &=& \int \!\! d^{3}x \Biggl\{ 
%      -\frac{1}{4g^2} {\rm tr}F_{\mu\nu}F^{\mu\nu}
%      - \Phi_i^{\dag} \left( D_\mu D^\mu + m^2_i \right) \Phi_i
%      + \overline{\Psi}_a\left( iD\!\!\!\! / - m_a \right)\Psi_a
%      + {\cal L}_{GF}
%      \nonumber\\
%  && \qquad\qquad\quad + G(A_\mu,\Phi,\Psi) \Biggr\}\ ,
%\label{action}
%\end{eqnarray}
\begin{equation}
S = \int \!\! d^{3}x \Biggl\{ 
      -\frac{1}{4g^2} {\rm tr}F_{\mu\nu}F^{\mu\nu}
      - \Phi_i^{\dag} \left( D_\mu D^\mu + m^2_i \right) \Phi_i
      + \overline{\Psi}_a\left( iD\!\!\!\! / - m_a \right)\Psi_a
      + G(A_\mu,\Phi,\Psi) \Biggr\}\ ,
\label{action}
\end{equation}
where
%
%${\cal L}_{GF}$ denotes gauge-fixing and ghost terms, and
%
$G$ is a general gauge-invariant function of 
$A_\mu$, $\Phi_i$ and $\Psi_a$.\footnote{
We assume that $G$ contains no quadratic terms in the fields.}
 We here assume that there is no bare Chern-Simons coupling.  
A generalization to
the theory with a non-vanishing bare Chern-Simons coupling will be 
discussed in the next section.

It should be emphasized that we take the normalization
that the gauge coupling $g$ does not appear in the
covariant derivative $D_\mu$, 
i.e. $D_\mu \equiv \partial_\mu - i A_\mu$. 
It is known that in
supersymmetric gauge theories this normalization of the gauge
coupling preserves holomorphy and is crucial to prove the one-loop
exactness of the gauge coupling renormalization \cite{shifman-vainstein},
as stressed in Ref.\cite{murayama}.  
This normalization of the gauge coupling
turns out to be crucial also in our proof of 
the one-loop exactness of the Chern-Simons coupling renormalization,
as we will see below.

We shall give a proof of the non-renormalization theorem of the 
Chern-Simons coupling in terms of the {\it Wilsonian} effective action
(for a review, see \cite{review-of-wilson}), 
which is identical to the 1PI
effective action when there are no interacting massless particles.
In supersymmetric theories, it is crucial to distinguish two effective
actions, as pointed out by Shifman and Vainshtein 
\cite{shifman-vainstein}.
When interacting massless particles are present,
the 1PI effective action suffers from infrared ambiguities and
might suffer from holomorphic anomalies, which would lead to the violation
of non-renormalization theorems 
\cite{holomorphic-anomaly1,holomorphic-anomaly2,holomorphic-anomaly3,
holomorphic-anomaly4}.  
These are absent in the Wilsonian effective action.  
Interestingly, a similar 
situation happens in our problem.  It has been reported, in 
Refs.\cite{b=0-1,infrared-singularity1,infrared-singularity2,
infrared-singularity3}, 
that non-vanishing radiative corrections to the
Chern-Simons coupling will appear in higher loops 
when massless particles are present.  
These higher order corrections are caused by infrared singularities.
To avoid such infrared problems, it should be
understood that our proof is applied to the Chern-Simons coupling
defined in the  Wilsonian effective action.  
Another advantage of the use
of the Wilsonian effective action is that non-renormalizable terms
in the action (\ref{action}) can be managed 
because high-momentum modes
beyond a cutoff are not included in the functional integration.
The momentum cutoff, however, gets into trouble with gauge invariance
\cite{gauge-vs-cutoff}. 
This subject is beyond the scope of this letter
and we assume that the cutoff respects gauge 
invariance in the Wilsonian effective action.

Our proof of the non-renormalization of the Chern-Simons term is an
application of the works by Dine \cite{dine} 
and Weinberg \cite{weinberg},
in which a simple proof of the non-renormalization \cite{fnprs} of the 
Fayet-Iliopoulos $U(1) D$-term \cite{f-i} has been given 
in general supersymmetric gauge theories.  
The key technical tool used in their
proof is the Seiberg trick \cite{holomorphy} of regarding
coupling parameters as  the scalar components of external superfields.
Following the approach by Dine and Weinberg, we regard all
the couplings in the action (\ref{action}) 
as the external scalar fields.\footnote{
Precisely speaking, we regard the gauge coupling and all coefficients
of interaction terms in $G$ as the external scalar fields.
The mass parameters remain to be constants.}
Suppose that the Chern-Simons term 
(\ref{cs-term}) is generated in the Wilsonian effective action.  
Then, the induced Chern-Simons coupling 
$\kappa$ would depend on
the gauge coupling and other couplings appearing as coefficients of 
interaction terms in $G$.
However, if we regard the couplings as the scalar fields, 
the expression (\ref{cs-term})  is {\it not}
gauge-invariant even under {\it small}
gauge transformations unless $\kappa$ is a constant,
i.e. independent of all coupling parameters.
The graphs independent of all couplings are
just the one-loop ones generated from 
the action without $G$ in Eq.(\ref{action}).
%This can be seen as follow:
%Suppose that the Chern-Simons term is renormalized
%by a $L$-loop graph (with two or three external gauge field lines).
%Since $\kappa$ must be independent of all the coupling parameters
%appearing in $G$, $G$ can have no effects on the renormalization
%of $\kappa$.
%With $G$ ignored, perturbation theory based on the action
%(\ref{action}) gives the relation $L={\cal N}_{g^2}+1$,
%where ${\cal N}_{g^2}$ is the power of $g^2$ of the
%$L$-loop graph.
%Since $\kappa$ must also be independent of $g$, ${\cal N}_{g^2}$
%should be equal to zero.
%Therefore, just the one-loop graphs generated from
%the action without $G$ in Eq.(\ref{action})
%can make corrections to $\kappa$.
%This completes the proof.
%
%
%
\section{Remarks}

We have presented a very simple (rather trivial) proof of the 
non-renormalization theorem of the Chern-Simons coupling in 
$2+1$-dimensional gauge theories 
without a bare Chern-Simons coupling.
Most of the results given in this letter have already been known but
some of them are new: For abelian theories, our proof can apply
to a wider class of gauge-invariant theories than those considered by 
Coleman-Hill \cite{coleman-hill}, in which gauge interactions have been 
restricted to the minimal coupling.  
For nonabelian theories, our results are new and
our proof has revealed that only {\it small} gauge 
invariance is enough to ensure the absence of higher order
corrections.  We do not need to require large gauge invariance at all.

We have started with general
gauge theories without a bare Chern-Simons coupling.
When a bare Chern-Simons term is present, our proof could not
apply to this case.
This is because the Chern-Simons term (\ref{cs-term}) would not be
gauge-invariant if the bare Chern-Simons coupling is regarded as
a scalar field.
For abelian theories, this problem can be remedied
by still keeping it a constant and by absorbing it into
the gauge field propagator.
This does not, however, work for nonabelian theories
because the nonabelian Chern-Simons term
includes an interaction term.
Although we do not know 
whether our proof can be generalized to this case, 
it may not be unreasonable to expect that a similar proof could
work for Yang-Mills-Chern-Simons gauge theories 
because no higher order corrections to the Chern-Simons coupling 
have been proved for
pure nonabelian Chern-Simons gauge theories 
\cite{pure-cs1,pure-cs2,pure-cs3}.

As mentioned in Section 2, it is crucial, in our proof, to take the
convention that the gauge coupling does not appear
in the covariant derivative.
One might expect that the proof given in the previous section could hold
in other conventions of the gauge coupling, for example, the canonical 
normalization, 
in which $A_\mu$ in Eq.(2) may be replaced by $gA_\mu$.
This is not, however, the case
because in the canonical normalization
the gauge kinetic term would 
%not be gauge-invariant 
break gauge invariance
if we think of the gauge coupling as a scalar field.  
Furthermore, the rescaling of
the field, $A_\mu \rightarrow gA_\mu$, 
will cause another problem that the functional measure
may not be invariant under the rescaling and
a Chern-Simons term could be generated from the
integration measure \cite{murayama}.  It would be of interest to clarify
the relation between different conventions of the gauge coupling.

It may be instructive to make a comment that our proof tells us that
the Chern-Simons term (\ref{cs-term}) 
can be generated at most at one-loop but does
not forbid the appearance of \lq\lq would be" Chern-Simons terms 
in the effective action \cite{higgs1,higgs2,higgs3}.  
For example, a term 
$\epsilon^{\mu\nu\lambda}\Phi^{\dag}F_{\mu\nu}D_{\lambda}\Phi$, 
which is manifestly
gauge-invariant and may be generated at higher loops, may reduce to a
Chern-Simons term after replacing $\Phi$ 
by its expectation value $<\Phi>$ in
the Higgs phase.

The final remark is as follows: In the previous section, we have insisted
that the induced Chern-Simons coupling must be independent of all
coupling parameters.  This will be true even non-perturbatively as long 
as gauge symmetry is preserved in the Wilsonian effective action.
We have then concluded, from this fact, 
that the radiative corrections to 
the Chern-Simons coupling come from only the one-loop diagrams 
generated from the action with $G=0$ in Eq.(\ref{action}).  
This conclusion is
true as long as the couplings are treated perturbatively
but is this true even non-perturbatively?  
At the moment, we have no definite answer to this question.  
This is because the requirement of gauge
invariance would not forbid
the induced Chern-Simons coupling, for example, to be a function of 
$\lambda/|\lambda |$ 
for some coupling parameter $\lambda$, since
$\lambda/|\lambda |$ looks like a \lq\lq constant" 
as long as $\lambda$ is restricted to be positive
(or negative).
In fact, such dependence is known to appear at one-loop
for $\lambda =$ the fermion masses and the bare Chern-Simons
coupling.
Since the mass parameters are incorporated into the propagators
and are not treated perturbatively, this may suggest that
a locally-constant function of the couplings, which does not
contradict gauge invariance, could appear non-perturbatively.
%In supersymmetric theories, holomorphy can avoid such dependence
%\cite{dine,weinberg}, but in non-supersymmetric theories
%gauge invariance alone seems insufficient to avoid it.
It would be of great interest to generalize the perturbative 
non-renormalization theorem to non-perturbative one.
%
%
%
%
% ACKNOWLEDGEMENT
\subsection*{Acknowledgement}

We would like to thank for T. Matsuyama and H. So
for useful comments and discussions.

\end{document}